\documentstyle{article} 
  
\begin{document}
\large
 ~

\vspace{7.5cm}

 \centerline{\huge \bf PKS 1932-464:~}

\vspace{0.5cm}

\centerline{\huge \bf a jet-cloud interaction}

\vspace{0.5cm}

\centerline{\huge \bf in a radio galaxy?}
 
\vspace{2.5cm}

M. Villar-Mart\'\i n$^1$, C. Tadhunter$^1$, R. Morganti$^{2,3}$, N. Clark$^4$,
 N. Killeen$^3$, 
 D. Axon$^4$
 
\vspace{0.5cm}

\small{
$^{1}$\,Dept. of Physics, University of
Sheffield, Sheffield S3~7RH, UK
\vspace{0.2cm}

$^{2}$\,Istituto di Radioastronomia, Via Gobetti 101,  40129 Bologna, Italy
\vspace{0.2cm}

$^{3}$\,CSIRO-ATNF, PO Box 76, Epping NSW 2121, Australia
\vspace{0.2cm}

$^{4}$\,Space Telescope Science Institute, 3700 San Martin Drive, 
Baltimore MD21218, USA } 
 
\vspace{0.5cm}
\large

{\it Accepted for publication in Astronomy \& Astrophysics}

\newpage
~
 
\vspace{1cm}
 
{\huge \bf Abstract}
 
\vspace{1cm}
 
We present optical and radio images, and long-slit spectra of the radio
galaxy PKS~1932-464 (z=0.230).  Our main goal is to determine whether the
observed pro\-perties of the extended emission line nebulosity in this
object show evidence for interactions between the radio jet and the
extended ionized gas, or whether they can be explained in terms
of the illumination of the extended gas by the active nucleus.  Although
the data do not show compelling evidence for jet-cloud interactions, the
existence of two well distinguished emitting line regions with very
different kinematics and ionization levels suggests the presence of such
interactions.  The large-scale gas distribution is complex and suggests
a gravitational interaction between the host galaxy of PKS~1932-464 and a nearby companion
galaxy.  The detection of a broad H$\alpha$ emission line in the nuclear
spectrum of the object provides evidence that this object is a broad
line radio galaxy (BLRG) rather than a narrow line radio galaxy (NLRG)
as pre\-viously supposed. 
 
\newpage

\section{Introduction}

	The general properties of the extended gas of most power\-ful radio
galaxies at low redshift can be explained in terms of the illumination of the
ambient gas by the active nucleus.  However, there are some low and
intermediate redshift radio galaxies (z$<$0.5) in which both the kinematics and the
morphology of the extended emission line regions (EELR) provide strong evidence
for powerful interactions between the radio jet and the extended gas (Clark 1996, Clark
{\it et al.} 1996,1997a,b). This is  particularly evident
in high redshift ($z>$1) radio galaxies, which present collimated structures closely aligned with
the radio axis (Chambers {\it et al.} 1987, McCarthy {\it et al.} 1987) and
highly perturbed kinematics ({\it e.g.} Mc.Carthy {\it et al.} 1996,
van Ojik {\it et al.} 1996, Pentericci {\it et al.} 1997). 

We are carrying out a spectroscopic study of a small sample of powerful radio
galaxies at intermediate and low redshift  to un\-derstand the processes associated with the interactions
between the radio jets and the ambient gas, and to gauge the importance of
shocks in determining the observed properties of these objects.  The ultimate
goal is to determine the relative importance of the jet-cloud interac\-tion
phenomenon in the general population of power\-ful radio galaxies. 

In the first stage of this work we have carried out detailed spectros\-copic and
imaging observations of objects which were known from previous studies to be
strong candidates for jet-cloud interactions (Clark 1996, Clark {\it et al.}
1996,1997a,b).  The results provided compelling evidence for the strong influence
that the shocks can have on the kinematics, morphology and physical properties
of the gas. 

As a continuation of this project, we have selected the radio source
PKS~1932-464 ($z$=0.230) (J2000 RA: 19 35 56.6 $\delta$: -46 20 41.8), which has
an early-type host galaxy.  Initially, this object was observed as part of the Tadhunter {\it et
al.} (1993) spectroscopic survey of southern radio galaxies with radio flux
$S_{2.7GHz} >$ 2Jy.  These early spectroscopic observations revealed an
extensive emission line ne\-bulosity extending out to a radius of 23    arcsec
($\sim$113 kpc\footnote{$H_0 = 50$ km s$^{-1}$ Mpc$^{-1}$
and $q_0 = 0.0$ assumed throughout.}) along the radio axis.  This marked PKS~1932-464 as an
interesting object for future study, although the {\it a prio\-ri} case for a
jet-cloud interaction in this object was not as strong as the objects discussed
in Clark {\it et al.} (1996, 1997), we were simply aware that the
emission lines are extended along the radio axis.  We present below new optical
spectroscopic, and optical and radio imaging observations which we use to
disentangle which properties of the object are a consequence of AGN
illumination, and which are due to jet-cloud interactions. 

\section{Observations}

\subsection{Optical Imaging}
 
The optical images were obtained using the ESO Faint Object Spectrograph
and Camera EFOSC 1 on the 3.6m telescope at La
Silla Observatory, Chile, on the night of 12/7/94. The detector  is a Tek CCD with
 512$\times$512 pixels$^2$ of 27  $\mu$m$^2$ giving an image field size 
of 5.2' $\times$  5.2'. The projected pixel size is 0.61    arcsec.
The observations were carried out  in moderate 
seeing conditions (1.8    arcsec FWHM). 

	Images with two different filters were obtained. Filter \#626
 was used to obtain the [OII]+continuum images ($\lambda_0$=4586 \AA\ $\Delta\lambda$=109 \AA\
), while the continuum images were taken with filter \#718 ($\lambda_0$=5445
\AA\ $\Delta\lambda$=175 \AA\ ).  Two 300 second exposures were obtained
with the pure continuum filter and two 600 second exposures with the
line+continuum filter.  These two frames were combined afterwards to increase
the signal-to-noise ratio.

	To improve the resolution and reveal faint structures, the
images were deconvolved using Lucy-Richardson's algorithm in the STSDAS package
in IRAF.  A star in the same frame was used as a PSF and the number of
iterations needed to obtain similar spatial reso\-lution in both the
[OII]+continuum (10 iterations) and the pure continuum image (14 iterations). 
 Once the LR algorithm was applied, the
effective seeing (FWHM) after restoration is 1.1    arcsec.

 \subsection{Radio observations}

PKS~1932$-$464 was observed with the Australia Telescope Compact Array (ATCA)
on 1994 September 4, using a 6-km array confi\-guration and the standard
continuum correlator setup providing a bandwidth of 128~MHz with 16
independent 8-MHz channels.  The ATCA allows observations at two simultaneous
frequencies (1.3/2.3 GHz or 5/8 GHz) and to switch very rapidly between the two
set of frequencies during the observations.  We took advantage of this
facility and we observed PKS~1932--464 in all four different frequencies (i.e. 
1.3, 2.3, 5.8 and 8.6 GHz).  At each frequency the source was observed for 6h
in total, spread out in equal length cuts over 12h.  Observing at 8.6~GHz
enabled us to achieve a resolution of $\sim$1 arcsec, the highest resolution
currently possible with ATCA.

The primary flux density calibrator (1934$-$638) was observed at the beginning
and the secondary calibrator (1933--400) was observed every 30 min to track the
complex antenna gains with time.  The data were analyzed with the MIRIAD
package (Sault, Teuben \& Wright 1995).  All images were produced with standard
synthesis ima\-ging, deconvolution and self-calibration methods.  Additionally,
all our images were made with uniform weighting, which provides a better
synthesised beam (narrower main peak and smaller sidelobes) at the expense of
the loss of some sensitivity.

Together with the total intensity $I$,
images of the Stokes parame\-ters $Q$ and $U$ were also produced and from these
we obtained the polarized intensity image ($P=(Q^2+U^2)^{1/2}$) and
position-angle ima\-ge ($\chi = 0.5 \arctan (U/Q)$).  The rms noise of the
signal-free portion of the $I$  and $P$ images are given in Table 1.  The
polarized intensity, and, as a consequence, the fractional polarization
($m=P/I$) were estimated only for the pixels for which $P>5\sigma_{QU}$.

Because all the observations were made with the
same configuration, the beam size is different for each frequency.  This makes
it difficult to use the four frequencies for a study of spectral index,
depolarization and rotation measure (RM).
The beam sizes at different frequencies are given in Table~1.

\begin{table}[ht]
\centering
\caption{Radio parameters.}
\begin{tabular}{lccc}
\hline
Freq. & Beam size          & $\sigma_I$         & $\sigma_P$  \\
      &  arcsec (degrees)  &  mJy beam$^{-1}$   &   mJy beam$^{-1}$ \\
\hline
 1.3GHz    & 9.9x3.9 (p.a. 15)  &   4.0      &    1.0                 \\
 2.3GHz    & 5.9x2.4 (p.a. 16)  &   1.4      &    0.6                 \\
 5.8GHz    & 3.0x1.2 (p.a. 19)  &   1.5      &    0.2                 \\
 8.6GHz    & 2.0x0.8 (p.a. 19)  &   1.4      &    0.3                 \\
\hline
\end{tabular}
\end{table}

\subsection{Long-slit spectroscopy}

The spectroscopic observations were carried out on the nights 27-28/9/94 
using   the Royal 
Greenwich Observatory (RGO) spectrograph 
on the Anglo Australian Telescope. The detector is a  Tek CCD with 226$\times$1024 pixels of 27 $\mu$m$^2$, resulting in a
spatial scale of 0.81 arcsec per pixel.  The  1.5    arcsec slit was oriented
at  PA 270.

\begin{table*}[hb]
\centering
\caption{Log of the spectroscopic observing run.}
\small
\begin{tabular}{lllllll} 
\hline
 Central $\lambda$ &  Grating & Exp time  & Spectral range & Resol (\AA\ )  & Seeing (")	& Slit   (")\\	\hline
4600 	&  1200V  &  6000 & 4210-5030 & 1.7  & 0.6     &  1.5     \\ 	
5800 	&   600R &  6000 &  4990-6590 &   3.3  & 1.0    &  1.5    \\ 	
6164	&  1200V  &  5400 & 5760-6580 &  1.5 & 0.6   &  1.5  \\ 
8300 	&  600R  &  3600 & 7520-9130 & 3.3 & 1.0     & 1.5     \\ 	\hline	
\end{tabular}
\end{table*}

We used two  gratings with different spectral resolutions and at different
angles to select a wide spectral range.  The spectral dispersions obtained
were 1.57 \AA\ pixel$^{-1}$ with the 600R grating  and 0.80 \AA\ pixel$^{-1}$ 
with the 1200V grating. 

The reduction of the data was done using standard methods provided in IRAF. 
The spectra were bias subtracted and divided by a flat-field frame (dome
flat-field).  Illumination corrections along the slit were found to be
negligible.  Cosmic ray events were removed.  The spectra were calibrated in
wavelength using comparison spectra of a CuAr arc taken before and after each
object.  Sky lines were carefully subtracted and the spectra corrected for
atmospheric extinction with the aid of mean extinction coefficients for the AAO. 
Molecular bands due to atmos\-pheric absorption were removed
 separately from all the 
spectra where they were evident. The bands were modeled appropriately
for each frame, using the spectra of different standard stars taken with
the same slit width as the one used for the target.

         For each night, grating and grating angle, we built a mean res\-ponse
curve from the standard stars observed that night with the same filter and
wide slit.  Each target frame was flux calibrated with the corresponding
response curve.  The spectra were also corrected for Galactic reddening,
E(B-V)=0.04, value based on Burstein and Heiles (1984) maps, using the
empirical selective extinction function of Cardelli {\it et al.}
(1989).  

        Both IRAF and STARLINK (DIPSO) routines were used to measure the
emission line fluxes.  For the blends, decomposition procedures were used in
STARLINK (DIPSO), fitting several Gaussians at the expected positions of the
components.  As the number of mathe\-matical solutions is very large, we applied
theoretical constraints when necessary (like fixed ratios between line fluxes
or sepa\-ration in wavelength).  In this way, we constrainted the range of
solutions to a much smaller space where the resulting models have physical
meaning.  More details on the fitting procedures will be given in section
\S3.6. 

\section{Analysis and Results}

\subsection{The optical morphology.}
  
	The raw and reconstructed [OII]+continuum images show complicated structures of 
line emission, which
bear li\-ttle resemblance to conical shapes expected in the case of anisotropic
illumination of homogeneously distributed gas by a hidden AGN (see Fig.~1, 
top panels). The images show that the gas
is distributed in clumps and filaments. This is observed in many radio
galaxies and is likely to be due to the  original inhomogeneous distribution of
the gas in the early type host galaxy (Tadhunter 1990).

	The main body of the galaxy shows a double structure in [OII]. A dark
band defines two regions: one of them is coincident with the optical continuum
nucleus, while the other --- named ARM in Fig.~1, top left panel --- is
situated $\sim$5    arcsec ($\sim$25 kpc) to the NE of the optical continuum
nucleus.  The separation between these two components is clearer in the
bottom-right panel in Fig.~1. This double structure is not visible in the
continuum image (Fig.~1, bottom-left panel).

\begin{figure*}
\includegraphics{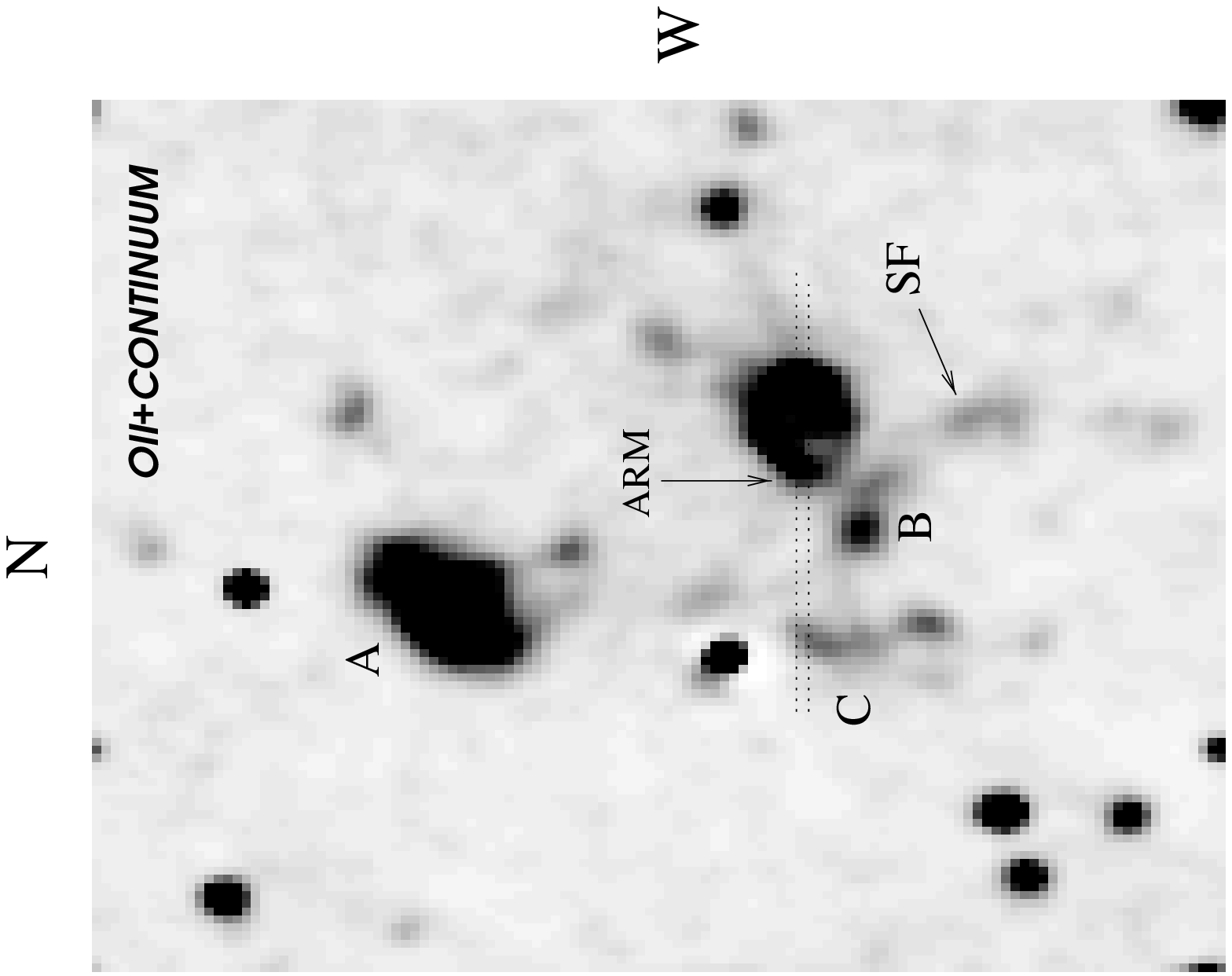}
\includegraphics{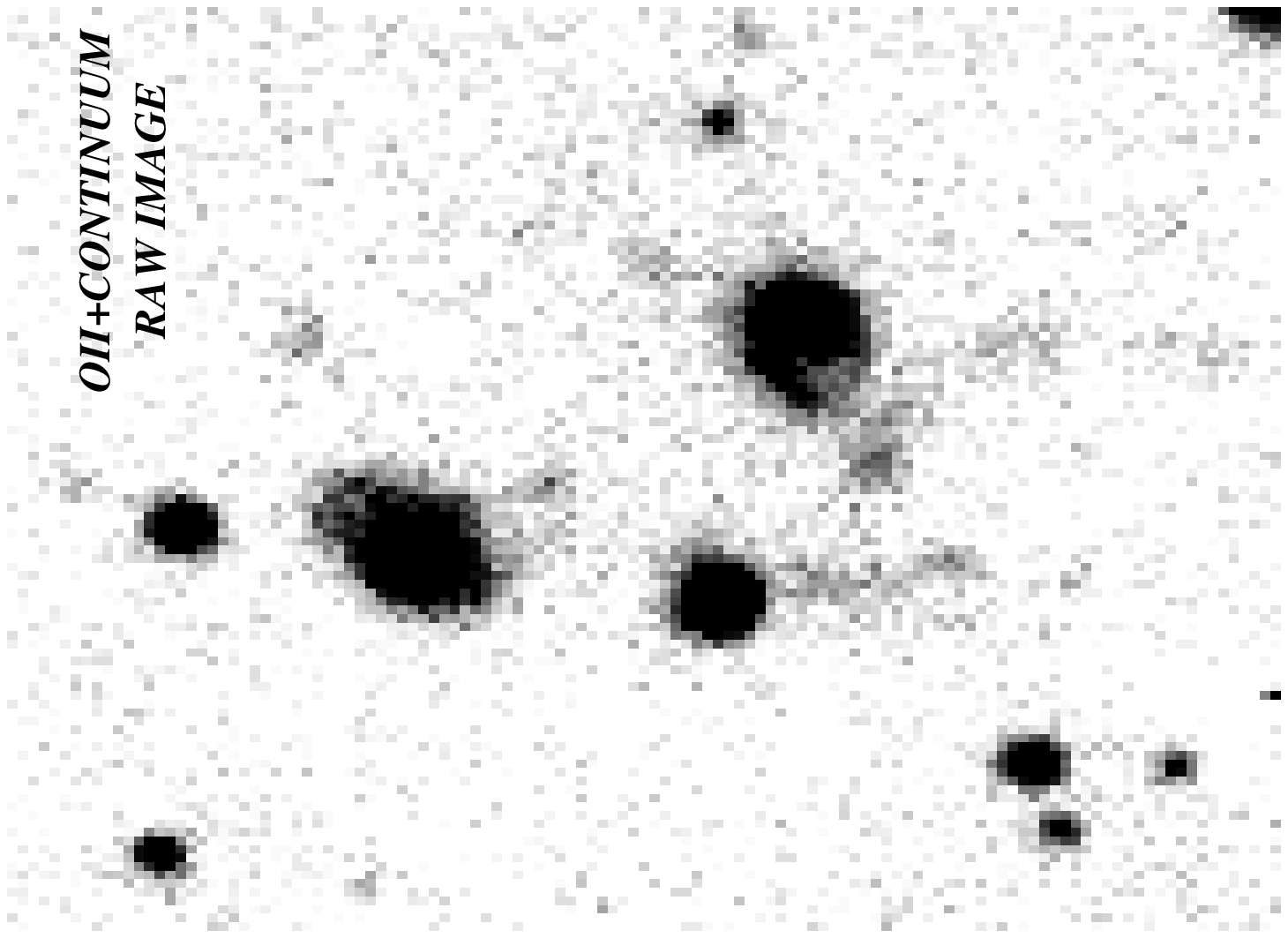}
\vspace{3.5in}
\includegraphics{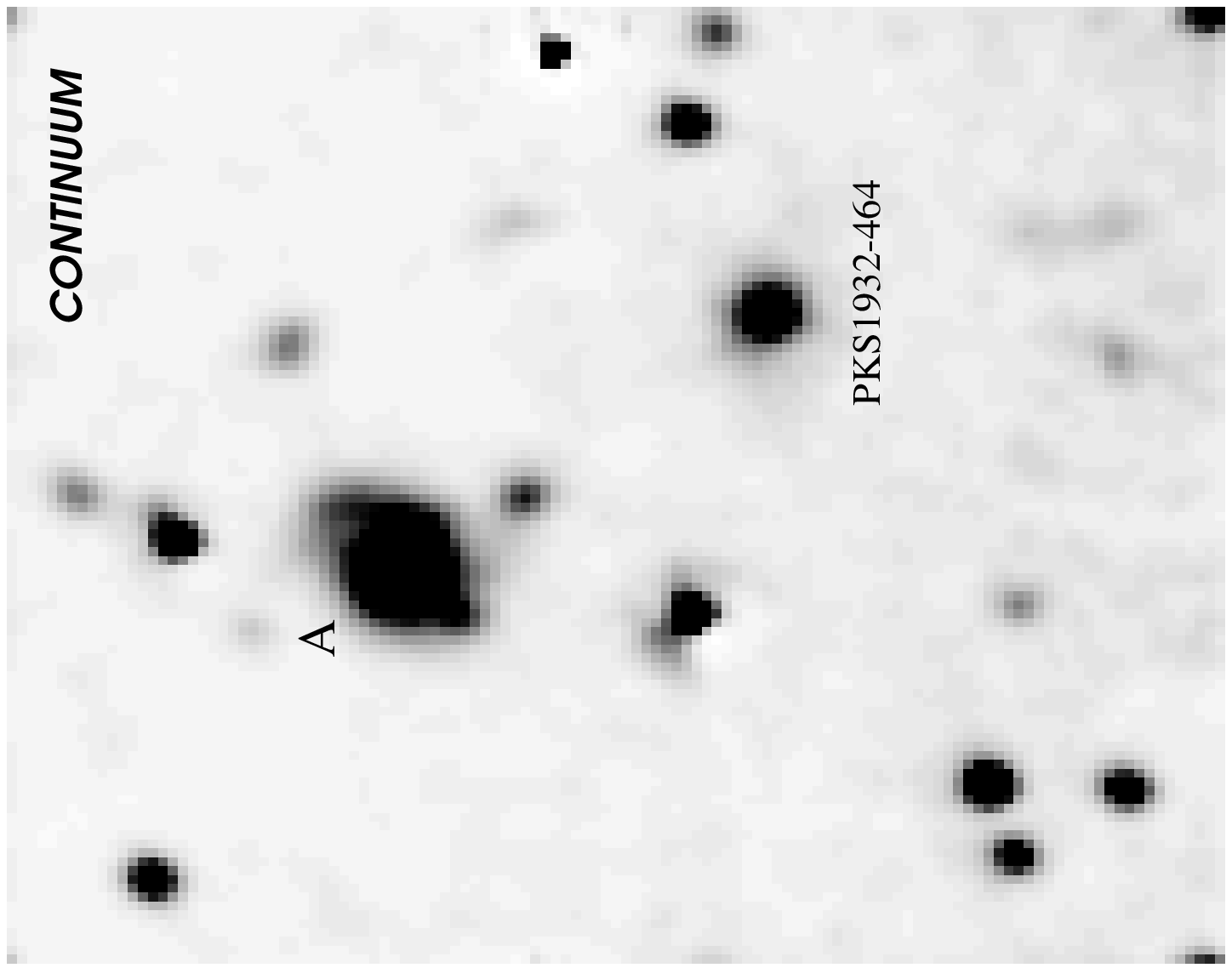}
\includegraphics{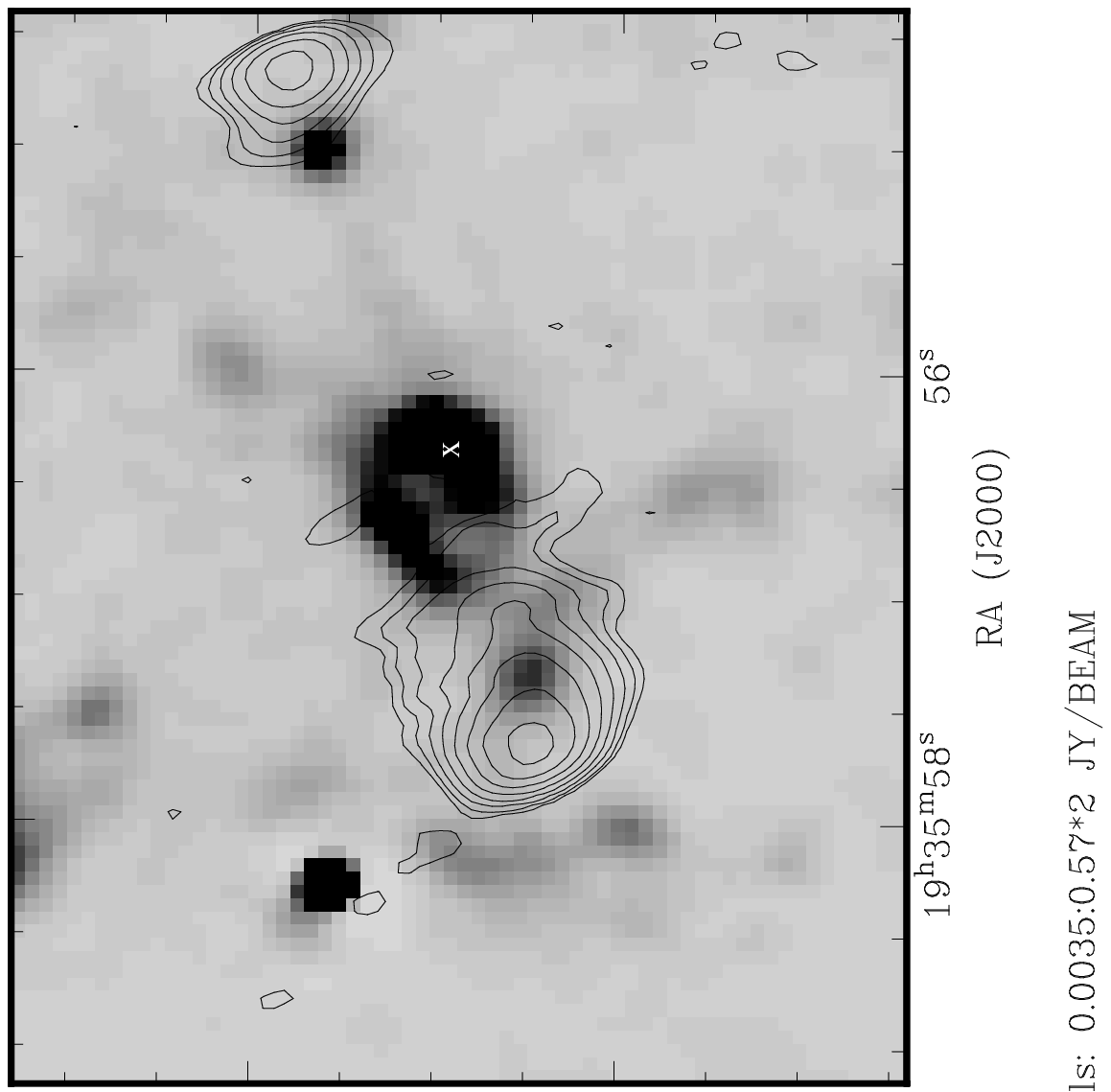}
\vspace{3.5in}
\caption{Top-left: [OII]$\lambda$3727+continuum image, deconvolved with Lucy's algorithm. Top-right: Raw image.
Most of the diffuse clumpy features surrounding the galaxy are pure line emitting regions. They do not appear
in the continuum image (bottom-left), which has been deconvolved to the same
spatial resolution. The slit position used for spectroscopy is shown in the top-left [OII]+continuum image. The pixel size in these images is 0.61
and the field is $\sim$65  arcsec E-W and 75     arcsec N-S. ~~~~~~~~~
Bottom-right: Radio contours (5.8 GHz) overplotted on the  [OII]+continuum image.
The field is $\sim$50  arcsec E-W and 45     arcsec N-S.
}

\end{figure*} 

 Structures of clumpy gas are spread around the main body of the galaxy.  An
interesting feature is the filament extending towards the south (named ``SF'',
for southern filament, in Fig.~1): it is a narrow, clumpy emission line
feature which extends to a maximum radial distance of $\sim$17    arcsec
($\sim$84 kpc).  We believe that this component lies at the same redshift as
the radio galaxy because it appears only in the [OII]+continuum images and not
in the continuum images.  The projected thickness of this arc is $\sim$6    arcsec  ($\sim$30 kpc).  The bright blob (feature ``B'' in Fig.~1) could be
connected with this southern filament. 

An inverted S-like or arc-like feature is seen to the E at a projected
distance of $\sim$19    arcsec ($\sim$ 94 kpc) (feature ``C'').  Although this
feature bears some resemblance to the bright emission line arc circumscribing
the radio lobe in PKS~2250-41 (Clark {\it et al.} 1997a), in this case the
filament lies significantly to the east of the eastern radio lobe 
(Fig.~1, bottom and see below). The astrometry was done using several stars
in the frame, for which the positions were accurately known. The astrometry
was mainly li\-mited by the error on the radio and optical reference system
($\sim$1 arcsec).

There is an interesting object in the field which is worth mentio\-ning.  The
galaxy named `A' in Fig.~1 could belong to the same system as PKS~1932-464,
but we do not know its redshift.  
Spectroscopy of the object will provide the answer.  The continuum image
su\-ggests the presence of two spiral arms, while the [OII]+continuum image
reveals more chaotic structures. 

\subsection{The radio structure}
                   
The radio images of PKS~1932--464 at two of the four observed frequencies are
shown in Fig.~2 (top).  The 8.6~GHz image represents the highest resolution
obtained --- the image at 5.8~GHz looks very similar to this --- while the
2.3~GHz image shows the morphology of the source at intermediate resolution
(the 1.3~GHz image is very similar).

\begin{figure}
\includegraphics{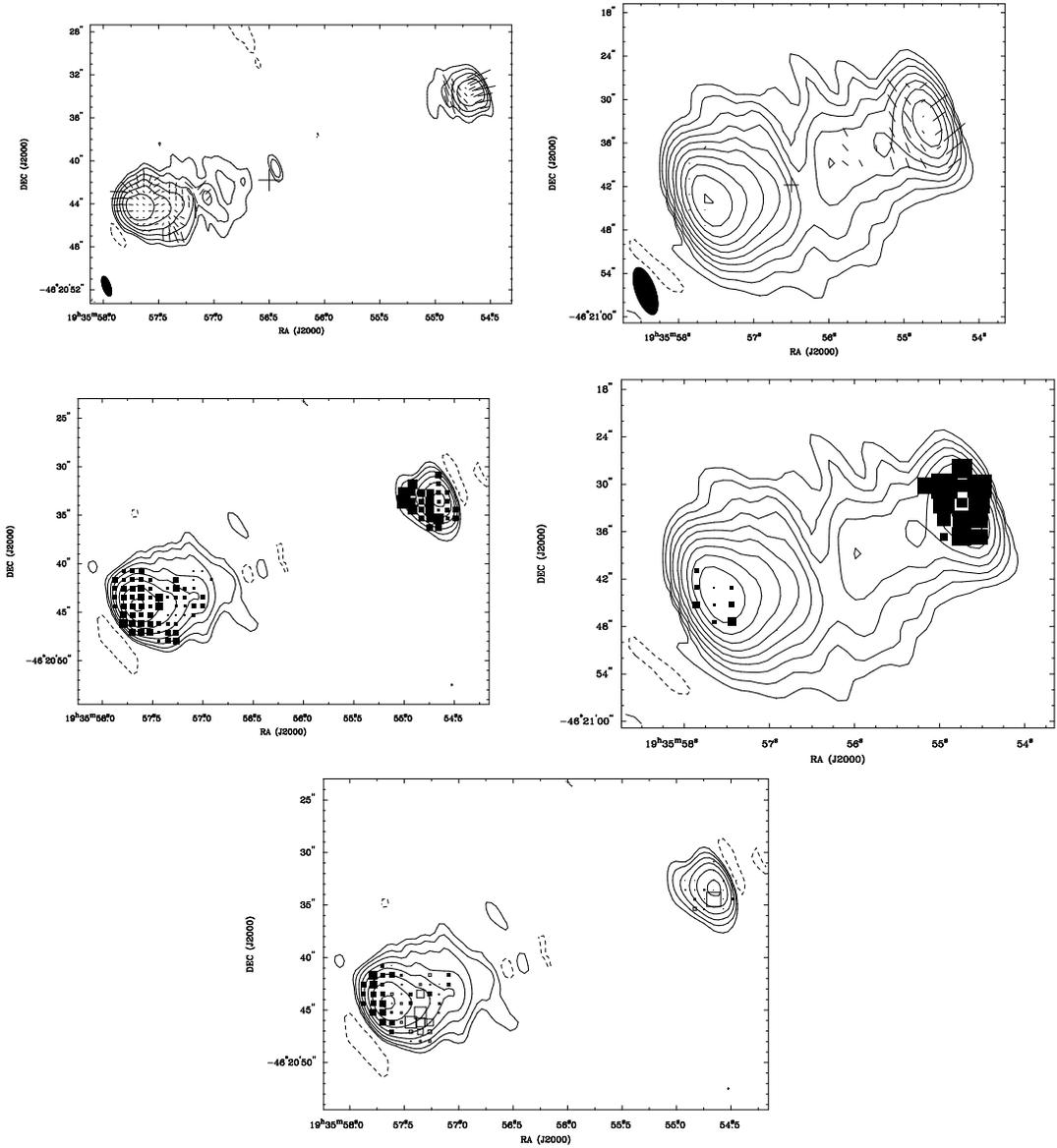}
\vspace{5in}
\caption{Top-left: Radio contours of the total intensity at 8.6~GHz with superposed
vectors whose length is proportional to the fractional polarization and whose
position angle is that of the electric field. The contour levels are -1, 1, 2,
4, 8, 16, 32, 64, 128, 256, 512, 1024 $\times$ 4 mJy beam$^{-1}$.
Top-right: Radio contours of the total intensity at 2.3~GHz with superposed 
electric field vectors. The contour levels are -1, 1, 2,
4, 8, 16, 32, 64, 128, 256, 512, 1024 $\times$ 4 mJy beam$^{-1}$.
Middle-left: depolarization between 5.8 and 8.6 GHz. Smaller squares represent
smaller $D$ values (stronger
 depolarization). Square sizes range between $D=$0.19 (small squares) and 2.4 (large
squares). Middle-right: depolarization between 2.3 and 5.8
GHz. Square sizes range between $D=$0.06 (small squares) and 1.6 (large
squares) Bottom: Faraday rotation between 5.8 and 8.6 GHz. Filled squares are
positive value of the rotation, open squares negative values.
}

\end{figure}

It is clear that the radio morphology of PKS~1932--464 is typical of
Fanaroff-Riley type II radio sources, with bright hot-spots in the lobes.  No
jet structure has been observed.  The two lobes are situa\-ted asymmetrically
compared to the optical nucleus (marked as a cross in Fig.~2).  The western
lobe is at a projected distance of $\sim$ 11 arcsec ($\sim 54$ kpc) from the
optical nucleus of the galaxy whilst the eastern lobe is $\sim 7$ arcsec
($\sim 35$ kpc) from the nucleus. 

At 8.6~GHz, the W side mainly shows the hot-spot that appears to be only
slightly resolved.  On the E side, together with the hot-spot, we can also see
low-brightness, extended structure reaching back towards the optical nucleus. 
At 2.3~GHz, also the W lobe shows an extension toward the optical nucleus.

At 8.6~GHz, we can distinguish a core component between the
lobes although it does appear to be offset (by $\sim 1.2$ arcsec) from the
position of the optical nucleus (see Fig.~2). Such an off-set is inside
the range we expect given the uncertainty on the astrometry of the
radio-optical frames. If we assume that
the unresolved radio component is the core, we get a radio flux density of
$S_{\rm core}^{\rm 8GHz} \sim 9$ mJy.  
 
Values for the flux
densities of the different components at diffe\-rent frequencies are given in
Table~3. 
 
\subsection{Radio polarization}

The values of the mean fractional  polarization $m$ in the lobes are given in
Table~3.  In Fig.~2 are shown the pola\-rization vectors, whose length is
proportional to the fractional polarization and whose position angle is that of
the electric field, superposed on the intensity contour maps at both 2.3 and
8.6 GHz. 

Assuming negligible Faraday rotation at 8.6~GHz (see below) the electric field
in PKS~1932--464, is distributed radially at the edge of the lobes (i.e.  the
projected magnetic field follows the edge of the lobe) as shown in Fig.~2. 
This kind of structure is commonly seen in powerful radio galaxies.
The structure of the projected magnetic field in the inner regions of the E
lobe is more complex; a sharp change in position angle is visible (both at 8.6
and 2.3 GHz) in the W lobe between the edge of the lobe and the regions closer
to the nucleus.  There is also weak evidence of a $90^\circ$  discontinuity in
the E lobe at the transition from lobe to ``bridge'' (i.e.  region closer to
the nucleus). 

Although some caution is required because of the di\-fferent beam sizes, Table~3
shows that the W lobe is relatively strongly polarized ($> 10\%$) at every
frequency, with a tendency for the fractional polarization to increase at high
frequencies (where the effects of the beam smearing are also smaller).  On the
other hand, the E lobe shows a significant polarization only at 5.8 and 8.6
GHz.  The percentage of polarization drops very sharply at the lower
frequencies.  This may be due to  beam smearing 
and therefore,
in order to reduce the effects of different beam size and investigate
polarization and spectral index, we have produced an 8.6~GHz image with
the same resolution as the 5.8~GHz image, and we have also carried out a
similar procedure at 5.8~GHz and 2.3~GHz.  We lowered the re\-solution of
the higher frequency by a Gaussian weighting function (taper) applied to
the visibilities.  The values of the fractional polarization obtained
from the images with ``degraded'' resolution are also given in Table~3.

\begin{table}
\centering
\caption{Radio parameters.}
\begin{tabular}{llllll}
\hline
Freq. & Total Flux  & \multicolumn{2}{c}{W lobe} & \multicolumn{2}{c}{E lobe}   \\
      &             &    I(Jy)    &   m(\%)      &   I(Jy)  &  m(\%)  \\
           \hline
1.3~GHz        &  12.21    &   2.11   & 10.8$\pm 1.2$  &   9.90 & ~0.9$\pm 1.3$                      \\
2.3~GHz        &  ~7.05    &   1.23   & 12.9$\pm 1.3$  &   5.66 & ~0.9$\pm 1.1$             \\
5.8~GHz$^a$    &  ~2.50    &   0.32   & 17.9$\pm 1.7$  &   2.11 & ~6.6$\pm 0.7$             \\
5.8~GHz$^b$    &   ~--~      &   0.38   & 14.7$\pm 1.5$  &   2.18 & ~4.2$\pm 0.5$             \\
8.6~GHz$^{c}$  &  ~1.60    &   0.24   & 19.3$\pm 2.5$  &   1.30 & 10.0$\pm 1.4$                \\
8.6~GHz$^{d}$  &   ~--~      &   0.24   &  18.1$\pm 1.8$  &   1.34 & ~8.6$\pm 1.0$             \\
\hline
\end{tabular}

$^a$ full resolution (3.0x2.4 arcsec beam size) \\
$^b$ same resolution as the 2.3~GHz\\
$^c$ full resolution (2.0x0.8 arcsec beam size) \\
$^d$ same resolution as the 5.8~GHz \\

\end{table}

Using the images with matched beam size we have estimated the depolarization
ratio between two frequencies, defined as $D=m_{\nu_1}/m_{\nu_2}$ with $\nu_1$
and $\nu_2$ the lower and higher frequency respectively.  We find
depolarization va\-lues of $0.79$ for the E lobe and $\sim 1$ (i.e.  no
depolarization) for the W lobe between 5.8 and 8.6~GHz.  Between 2.3 and 5.8
GHz we find a $D=0.23$ (strong depolarization) for the E lobe and $D=0.89 $ for the W
lobe.  This confirms the strong depolarization in the E lobe when we go to lower
frequencies.  Fig.2 (middle) shows the depolarization va\-lues superposed on the
5.8 and 2.3~GHz total-intensity contours.  The size of the boxes is
proportional to the value of the depolarization: big boxes represent value
close to 1 (i.e.  weak or no depolarization), small boxes represent va\-lues
close to zero (i.e.  strong depolarization). 

In summary, no strong depolarization is found between 5 and 8 GHz in
PKS~1932--464 and only a marginal assume\-try in the depolarization is observed
(at these frequencies) between the two lobes.  However, a strong
depolarization is present in the E lobe of PKS~1932--464 for lower
frequencies, i.e.  between 2.3 and 5.8~GHz.  This is diffe\-rent from the
results obtained for PKS~2250-41 (Clark {\it et al.} 1997a) where, even at high
frequencies, a large depolarization is observed in the radio lobe close to the
EELR. 
 
From our data, we can only estimate the Faraday rotation measure ($RM$) between
pairs of frequencies (those with the same reso\-lution).  The $RM$ is defined as
$\chi(\lambda^2) = \alpha + RM \lambda^2$, where $\alpha$ is the intrinsic
position angle and $\chi$ the apparent position angle at the $\lambda$ of the
observations.  Since the position angles are ambiguous by $\pi$, the $RM$
calculated from only two frequencies are ambiguous by $\pm n\pi$ rad m$^{-2}$. 

Between 5.8 and 8.6~GHz we find a small $RM$ in the W lobe.  The median $ RM$
is $\sim -9$ rad m$^{-2}$ but ranging between --14 rad m$^{-2}$ in the eastern
part to 35 rad m$^{-2}$ in the western side of this lobe.  In the E lobe the $RM$ shows a
wider range of values from $\sim 181$ rad m$^{-2}$ in the eastern side to $\sim
-107$ rad m$^{-2}$ in the western part.  The $RM$ between 2.3 and 5.8~GHz shows
similar values in the W lobe with $RM \sim 19$ rad m$^{-2}$ (indicating that
no $\pi$ ambiguity should be present) but it is too uncertain in the E lobe.

The depolarization and the $RM$ observed in radio galaxies are commonly
attributed to the effect of an inhomogeneous, unresolved, foreground screen,
which rotates the polarization position angle randomly across the 
synthesized beam (the
effective vector sum of the radiation is thus depolarized).  Indeed, by using
the formula in Clark et al.  (1997), we find that an $RM$ gradient of only 25
rad m$^{-2}$ arcsec$^{-1}$ is required to cause the observed depolari\-zation
ratio to fall to a value of $D\sim 0.24$ between 2.3 and 5.8 GHz (13 and 6
cm).  Thus, the observed depolarization seems to be compatible with an
unresolved RM fluctuation (caused by a Faraday screen anywhere along the line
of sight) across the synthesized beam causing the electric vector to rotate
across the beam. 

	Although the identity of the depolarizing medium is uncertain,
it is possible that at least some depolarization is associated with the 
warm emission line gas. This is suppor\-ted by the fact that the region
of highest depolarization in the eastern lobe is coincident with the position of knot B in the emission line image (see Fig.1).

Finally, we also computed an image of the spectral index ($\alpha$, defined as
$S_{\nu} \propto \nu^{\alpha}$), between 5.8 and 8.6~GHz.  We find  
 $\alpha = -1.30$ for the E lobe, $\alpha = -0.67$ for the W
lobe.

\subsection{The spatial distribution of the emission lines.}

	The position of the slit is shown in Fig.1 (top-left panel). The
slit -- aligned along PA 270 -- contains the optical nucleus but 
 is rotated with respect to the radio axis
so that it does not contain the main radio features (hot spots).

	We have examined the spatial distribution of the emission line fluxes
by extracting 1-dimensional spatial cuts from the 2-dimensional frame, adding
the pixels along the spectral direction which contain the line emi\-ssion, and
subtracting a spatial profile of the conti\-nuum obtained in a similar way. 
This procedure was ca\-rried out for [OII]$\lambda$3727, H$\beta$,
[OIII]$\lambda$5007, [OI]$\lambda$6300 and H$\alpha$+[NII]$\lambda$6583.

 The spatial distribution of the lines is shown in Fig.~3 for
[OII]$\lambda$3727, H$\beta$, [OIII]$\lambda$5007, [OI]$\lambda$6300 and H$\alpha$+[NII]$\lambda$6583.

 \begin{figure}
\includegraphics{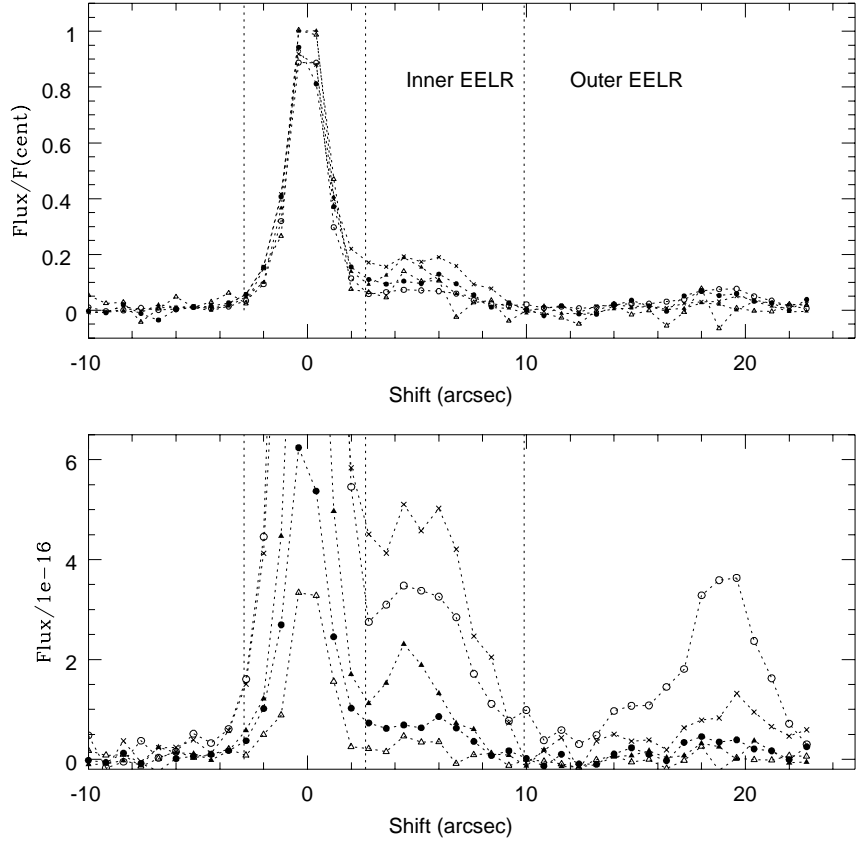}
\vspace{3.5in}
\caption[]{Integrated (in $\lambda$) spatial profiles of the lines 
 [OII]$\lambda$3727, [OIII]$\lambda$5007, [OI]$\lambda$6300, H$\beta$, 
H$\alpha$+[NII]$\lambda$6583. Top-panel: For every line, the flux has
been normalized to the value at the position of the continuum centroid. 
Bottom-Panel: The fluxes have been divided by 10$^{-16}$ to highlight
the differences in the spatial profiles of the different lines.
The three spatial
regions considered (see text) have been  separated by
vertical lines. Open circles - [OIII]$\lambda$5007; crosses - [OII]$\lambda$3727;
solid circles -  H$\beta$;
solid triangles -  H$\alpha$+[NII]; open triangles - [OI]$\lambda$6300.}
\end{figure}

The pure line spatial profiles reveal three main spatial regions: the nuclear
region (extending to radial distances of $\sim$3    arcsec or 15 kpc on both sides of
the nucleus), the {\it inner} EELR (extending to radial distances of $\sim$10.5 arcsec or 52
kpc to the East of the nucleus) and the {\it outer} EELR (extending to a maximum radial
distance of $\sim$23    arcsec or 113 kpc to the East of the nucleus).  The outer EELR coincides with the arc like feature mentioned in
section \S3.1 (feature ``C'' in Fig.~1), which lies outside the E radio lobe. 
The inner EELR corresponds to the extended gas near the main body of the
galaxy, also mentioned in section \S3.1 and named ``ARM'' in Fig.~1.  The
spatially integrated spectra for each region are shown in Fig.~4. 

The peak of the emission lies in the nuclear region for all the lines.  The
inner EELR emits stronger lines than the outer one, except for [OIII] which is
stronger in the arc.  [NII]+H$\alpha$ show a much steeper distribution than
the other lines, with the peak also closer to the nucleus. 

\begin{figure} 
\includegraphics{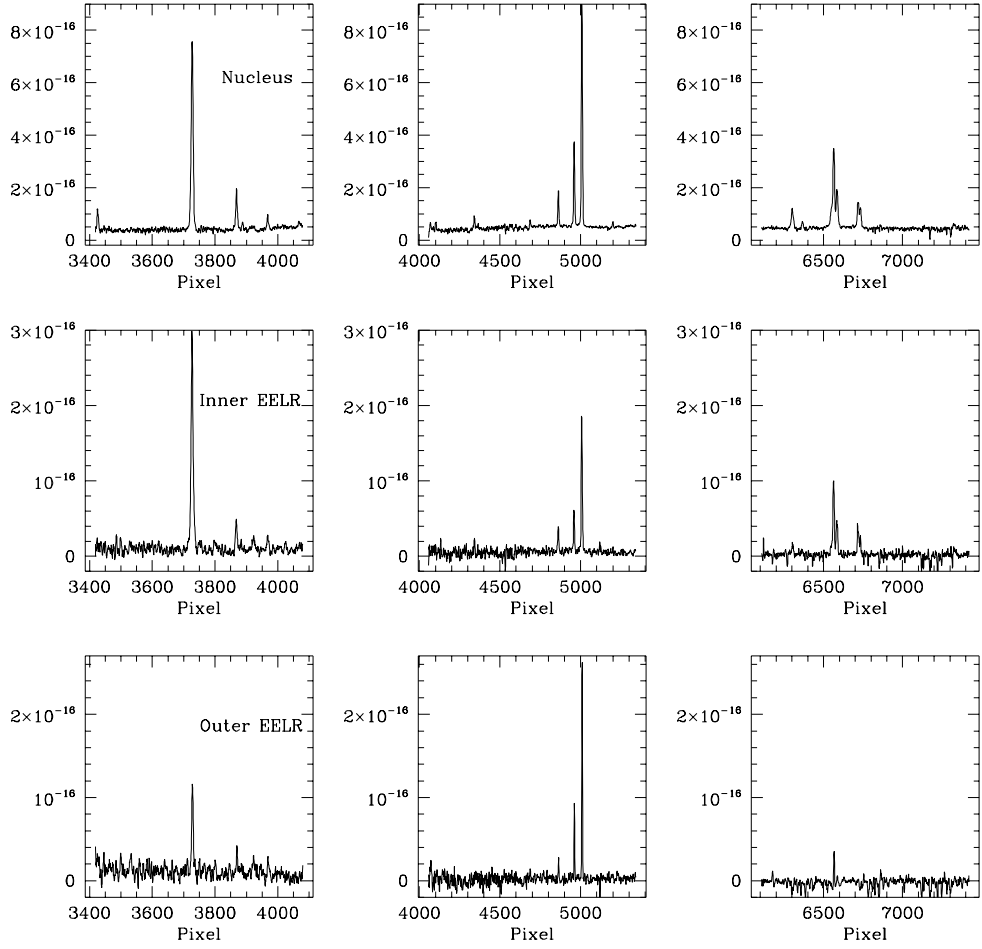}
\vspace{3.5in}
\caption[]{The spectra of the three spatial regions
revealed by the pure emission spatial profiles.  The upper panels show the
nuclear spectra.  The middle panels show the spectra of the inner EELR, which
corresponds to the extended gas, near the main body of the galaxy, named
``ARM'' in Fig.~1.  The bottom panels show the spectra of the outer EELR, which
coincided with the arc-like shape feature (named ``C'') in Fig.~1.  }
\end{figure}

\newpage 
\subsection{The line ratios and ionization gradient.}

 Table 4 shows the line fluxes (normalized to H$\beta$)
for some interesting lines. 
	The line ratios have not been corrected for reddening due to
 dust internal
to the galaxy PKS~1932-464. The highest measured Balmer decrements (3.6$\pm$0.4) indicate that  E$_{B-V}<$0.23, which is consistent with no significant reddening
within the errors. 

 \begin{table} 
\centering
\caption{Line fluxes relative to H$\beta$ for the three spatial regions considered in the text:
the nuclear region, the inner EELR and the outer EELR. H$\beta$ flux is given
in units of ergs s$^{-1}$cm$^{-2}$}
\begin{tabular}{llll} \hline
\hline
	&	Nucleus		& Inner EELR	& Outer EELR 	\\ \hline
~Flux(H$\beta$)	&  (1.17$\pm$0.04)$\times10^{-15}$	&  (2.9$\pm$0.3)$\times10^{-16}$   & (1.4$\pm$0.2)$\times10^{-16}$ \\  \hline	
~[NeV]$\lambda$3426 & 0.35$\pm$0.03	& $\leq$ 0.16	&  $\leq$0.21	\\
~[OII]$\lambda$3727 & 4.9$\pm$0.2 	&  7.6$\pm$0.9	&  4$\pm$1	\\
~[NeIII]$\lambda$3869  &  0.74$\pm$0.05	&  0.8$\pm$0.2	& 0.9$\pm$0.3	\\
~[Ne III]+H$\lambda$3967   &   0.22$\pm$0.03 	&  0.4$\pm$0.1	& 0.5$\pm$0.2	\\  
~H$\delta$ & 0.21$\pm$0.04	&  $\leq$ 0.16	&   $\leq$0.13	\\
~H$\gamma$ &  0.39$\pm$0.08	&  0.5$\pm$0.1	&  $\leq$0.30	\\
~[OIII]$\lambda$4363 & 0.09$\pm$0.03	& 0.2$\pm$0.1	&  $\leq$0.46	\\
~HeII$\lambda$4686 & 0.15$\pm$0.05	& 0.2$\pm$0.1
	&  0.4$\pm$0.2	\\
~H$\beta$	&  1.00	&	1.00	&	1.00	\\
~[OIII]$\lambda$5007 & 6.7$\pm$0.3	&   4.7$\pm$0.6	& 10$\pm$1	\\
~[NI]$\lambda$5199   & 0.17$\pm$0.04	&  0.2$\pm$0.1	&  $\leq$0.06	\\
~[OI]$\lambda$6300 & 1.01$\pm$0.08	&  0.6$\pm$0.2	& $\leq$1.03	\\
~H$\alpha$  & 3.3$\pm$0.2	&  3.6$\pm$0.4	&  1.8$\pm$0.6	\\
~[NII]$\lambda$6583  & 1.8$\pm$0.1	&  1.9$\pm$0.3	&  0.4$\pm$0.2	\\
~[SII]$\lambda$$\lambda$6716+6732  &  1.9$\pm$0.1  	& 2.3$\pm$0.5	&  $\leq$1.31	\\
\hline 
\end{tabular}
\end{table}

\begin{figure}
\includegraphics{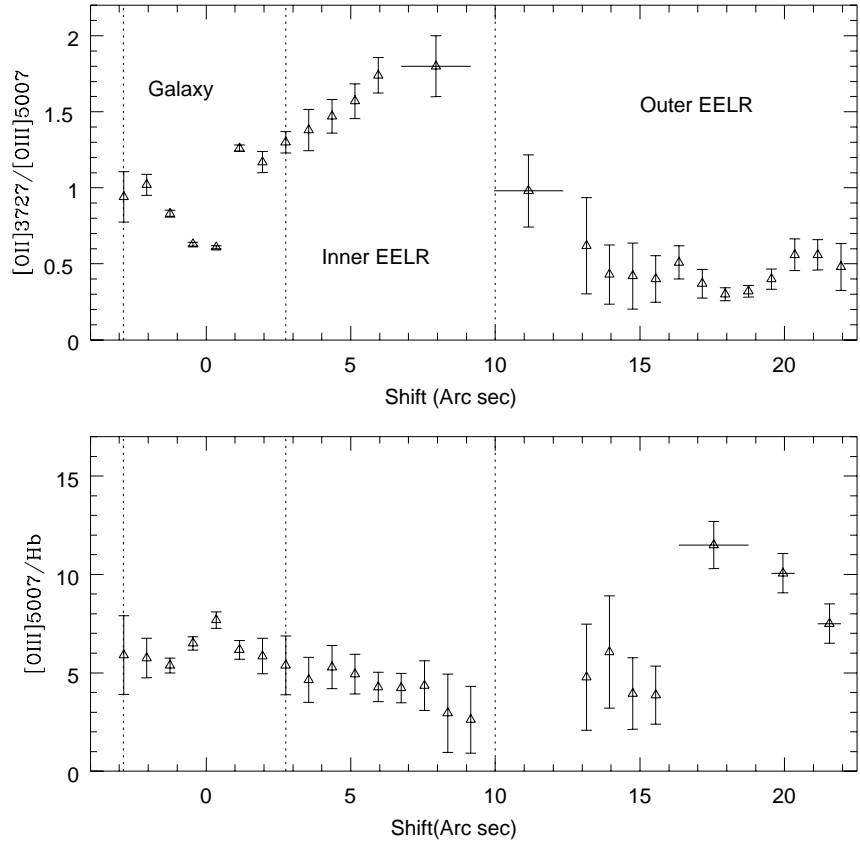}
\vspace{3.5in}
\caption[]{ The ionization gradient: Spatial variation of the [OII]$\lambda$3727/[OIII]$\lambda$5007 and [OIII]/H$\beta$ line
ratios.  Where the lines were too faint, several pixels were binned spatially.
This is shown by horizontal
lines in the plots, which indicate the pixels which were added.}
\end{figure}

The spatial variation of the ionization level of the gas 
is presented in Fig.~5, where we plot [OII]$\lambda$3727/[OIII]$\lambda$5007 and
[OIII]$\lambda$5007/H$\beta$ {\it vs.} projected distance from the nucleus.
 
The spectra containing the three main emission lines where observed on the same
night, with similar seeing and same slit width and position. Therefore, we
 are confident
that the [OII]/[OIII] measurements are reliable.  The agreement  with the
[OIII]/H$\beta$ ratio behaviour supports this point. 

We find  the 
ionization decreases  moving outwards
across the {\it inner} EELR (and E radio lobe), 
where it presents a minimum and then
rises  to  high values in the {\it outer} EELR.

\subsection{The kinematics}

	A close inspection of the 2-D frames  shows a clear 
diffe\-rence in the kinematics of the gas in the inner and the outer EELR
(see Fig.~6).
The    
inner EELR shows two distinct spectral components (A and B in Fig.~6), which are evident in all lines, while the  emission lines of the outer EELR are 
apparently simple.
The two spectral components in the inner EELR are spatially
separated: component A lies closer to the nucleus and is redshifted with
respect to component B. The peaks of the emission are 
separated by $\sim$2.4    arcsec  ($\sim$12 kpc) in projection.
Close inspection of the 2-D frame reveals a rather sharp cut between these
components in the spatial direction. This su\-ggest that we are observing two
spatially disconnected components, rather than a mixture of gases with
different kinematic properties.

\begin{figure} 
\includegraphics{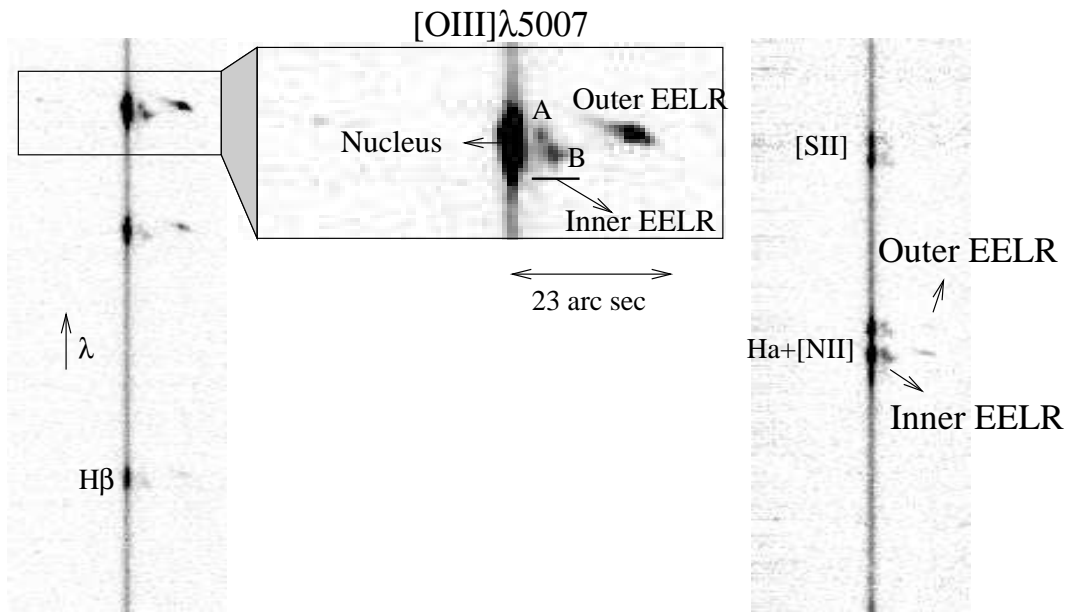}
\vspace{3.2in}
\caption[]{Left: 2-D spectrum showing [OIII]$\lambda$$\lambda$5007,4959 and H$\beta$ (red is up).
Middle: Zoomed [OIII]$\lambda$5007 region. Right: Red spectrum showing H$\alpha$+[NII] lines
and the [SII] doublet (top). The different kinematics in both EELR is evident:
the inner EELR shows split components while the outer one shows narrow simple
lines.}
\end{figure} 

 We have
studied the spatial variation of  the line widths  and 
the velocity shifts with
respect to the nuclear emission. 
We have also compared the behaviour of the different
lines for every spatial position.

	In order to do this, we isolated 1-D spectra from every spatial
pixel. We then fitted the line profiles with simple Gaussians, fixing constraints
 to obtain fits with physical meaning ({\it e.g.} the theoretical
flux ratios). The center of the Gaussian were taken as the central 
wavelength  of the line and the FWHM as the measured FWHM of the line. 
The measured line widths were  corrected for instrumental broadening.

	The results are shown in Fig.~7. Each symbol represents a 
diffe\-rent line.  All lines were resolved at every spatial pixel, except in a few
exceptional cases, which will be indicated by arrows (upper  li\-mits) in the plots.
Where the lines were too faint, several pixels were binned. This is indicated
by a thick dashed horizontal line,  showing the pixels which were added.
The main results are as follows.

\begin{figure}
\includegraphics{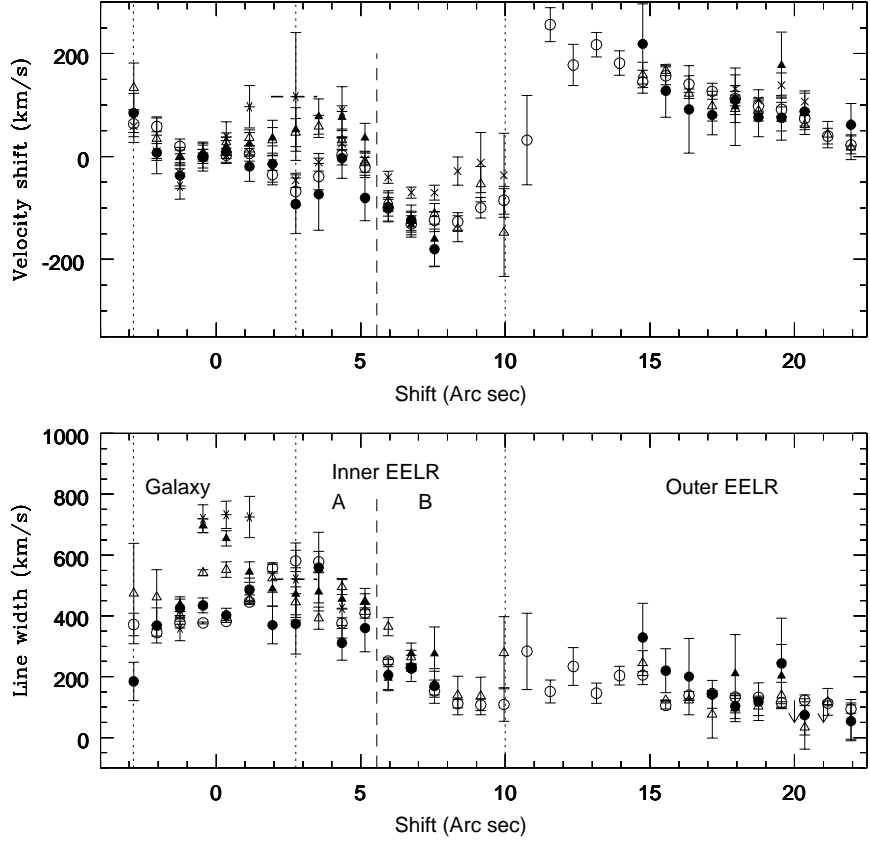}
\vspace{3.5in}
\caption[]{Variation in the radial velocity shifts (top) and line widths 
(bottom) along PA 270.
Open circles - OIII$\lambda$5007; crosses - [OII]$\lambda$3727;
solid circles -  H$\beta$;
open triangles -  H$\alpha$; solid triangles - [NII]6583; stars - [OI]$\lambda$6300. 
Line widths  for [OII] have not been plotted; due to the contribution of the
doublet components the measured widths could be misleading. [OI] line might
be contaminated by the [SIII]$\lambda$6312 line.
The three main spatial regions (see text) are
separated by dotted lines. The long dashed line marks the separation
between the two velocity  components A and B seen in the inner EELR.}
\end{figure} 
 
\vspace{0.2cm}\noindent
{\it The velocity field}
\vspace{0.2cm}

	The velocity field along PA270 (Fig.~7, top) 
shows no definite pattern and
the radial velocity range covered is $\sim$400 km s$^{-1}$, which is
comparable with the values measured in other radio galaxies with 
no signs of jet-cloud interactions ({\it e.g.} Tadhunter
{\it et al.} 1989, Baum {\it et al.} 1990). 

	The diagrams reveal that  the nuclear region, the inner EELR and the outer EELR are kinematically
distinct.

	 The inner EELR shows a velocity curve which is a continuation of the
velocity field in the nuclear region.  However, as expected from Figure 6,
there is a jump (clear in the [OIII] line) in the transition region between the
inner and outer EELR (at $\sim$10.5    arcsec from the nucleus).  Components
A and B of the inner EELR present different velocity shifts with respect
to the nucleus. At the spatial peak of intensity, B is redshifted  $\sim$150
km s$^{-1}$ with respect to the nucleus. Line emission from the peak of 
intensity of component A reveals a very small velocity shift in the line of sight
with respect to the nucleus.

	 The {\it outer} EELR shows velocity shifts which decrease outwards from
$\sim$250 km s$^{-1}$ to $\sim$25 km s$^{-1}$ in the region most distant  from the
nucleus.  This is also seen in the 2D frame (Fig.~6), where a tail of [OIII]
emission extends towards the nucleus. 

Both the inner and outer EELR show on average a similar velocity shift
($\sim$100 km s$^{-1}$) for the lines with respect to the nuclear region, but
with opposite signs.

\vspace{0.2cm}

{\it The line profiles}

\vspace{0.2cm}

	As a first approximation, we fit simple Gaussians for every line
and at every spatial position.	Large widths are measured for 
[OI]$\lambda$6300, [NII]$\lambda$6583 and H$\alpha$ in 
the nuclear regions
of the galaxy compared to [OIII]$\lambda$5007 and H$\beta$ in the same region. The large widths
measured for [NII] and H$\alpha$ could be due to the existence of an 
underlying broad H$\alpha$ component, which we have not taken into account.
On the other hand [OI]$\lambda$6300  could be contaminated 
with the line [SIII]$\lambda$6312. This is supported by the fact that the width of the [OI]$\lambda$6364 line is $\sim$480 km s$^{-1}$, similar to the values
measured for [OIII] $\lambda$5007 and H$\beta$.
The nuclear spectrum deserves a separated study and we will analyze it in
detail in the next section.

The {\it inner} EELR emits lines with split components
(see Fig.~6) which reveal two gaseous components moving in different ways: 
component A  (closer to the nucleus) shows similar line widths
($\sim$450 km s$^{-1}$) and radial velocity ($\sim$0 km s$^{-1}$) to the
nuclear region.   Component B is blueshifted by $\sim$150 km
s$^{-1}$ and shows narrower lines ($\sim$200 km s$^{-1}$).  This further
supports the idea that we are observing two kinematically distinct components in
the inner EELR.

The broader lines measured in  component A of the inner EELR
could be due to a contribution to the line profile from a spatial extension of
 component B towards the nucleus.  We have investigated
whether the line profile of component A can be fitted by
two narrow components, rather than a broad one.  The result indicates that the
best fit is, indeed, obtained with a single broader component.

	 The line widths are narrow in the outer EELR region (but still
resolved) compared to the inner EELR, showing $\sim$200 km s$^{-1}$ (FWHM), similar  to component B of the inner EELR (see Fig.~7).

Within the errors, all lines in the extended emission line regions show very similar widths and velocity shifts at
every spatial position.   	
 
\vspace{0.2cm}\noindent
{\it The nuclear spectrum}
\vspace{0.2cm}

A detailed analysis of the nuclear continuum spectrum by Dickson {\it et al.}
(1997) reveals a significant UV continuum excess compared with normal
early-type galaxies.  Possible origins
for the UV excess include: scattered AGN light, the light from young stellar
populations in the host galaxy, and direct AGN light.  However, the low
polarization measured for the UV continuum ($P < 1.6$\%, Dickson {\it et al.}
1997) rules out the idea that the UV continuum is dominated by a scattered AGN
component.  In this section we search for broad permitted lines in the nuclear
spectrum to gauge whether there is a significant contribution from direct,
rather than scattered, AGN light.

Attempts to fit the [NII] + H$\alpha$ blend using only narrow 
components  provide an unsatisfactory fit to the wings of the blend.
We find that the best fit to the blend requires a broad
component to H$\alpha$ with velocity width 2400$\pm$200km s$^{-1}$ shifted
by 70$\pm$80 km s$^{-1}$ with respect to the narrow component. The results
of the fit are presented in Table 5 and in Fig.~8. The detection
of broad H$\alpha$ indicates that
PKS~1932-464 is a broad line radio galaxy (BLRG), albeit with relatively 
weak broad lines.  Combined with the low UV polarization this result supports
the idea that there is a significant contribution from direct AGN light in 
this object.

  \begin{figure}
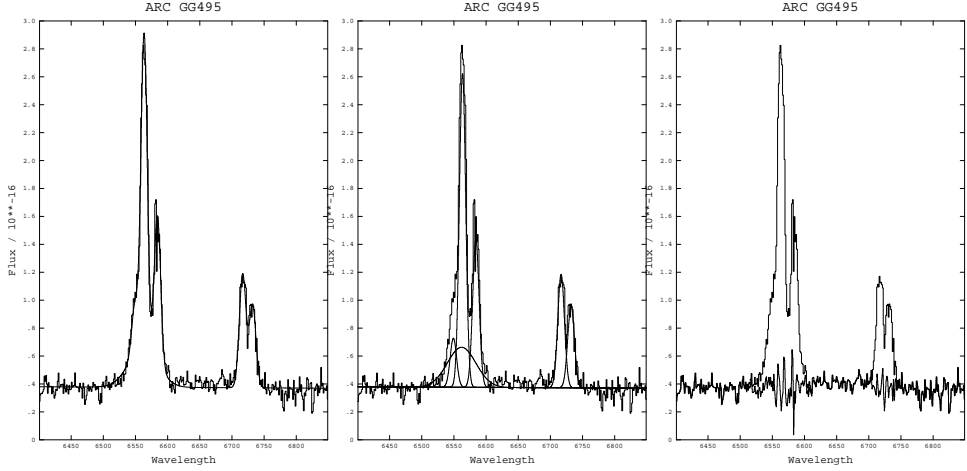

\includegraphics{f8a.eps}
\includegraphics{f8b.eps}
\includegraphics{f8c.eps}
\vspace{3in}
\caption[]{ Fit to H$\alpha$+[NII] and [SII].
Left: Data and fit. Middle: Data and individual components of the fit.
Rigth: Data and residuals.}
\end{figure} 

 The best fit to H$\beta$ line requires only a narrow component,  broad
wings are not detected. 
Taking into account the properties of the broad component of H$\alpha$,
we have checked whether the corresponding broad component to H$\beta$ should be
detectable.  Our predictions indicate that a broad component to H$\beta$ with the same velocity 
width as broad H$\alpha$ and a flux such that H$\alpha_{broad}$/H$\beta_{broad} \sim$3  would be undetectable.

\begin{table}
\centering
\caption{}
\begin{tabular}{llll} \hline
\hline
Line & Wavelength  & FWHM   & v$_{shift}$  H$\alpha_{B-N}$  \\ 
     &		   &	km s$^{-1}$  	&  	~~~~~ km s$^{-1}$	 	\\ \hline
~[NII]$\lambda$6543  &  6549.2$\pm$0.3   &  520$\pm$80  & \\ 
~H$\alpha_N$ &  6563.4$\pm$0.4 &  510$\pm$70  & \\ 
~[NII]$\lambda$6583  &  6584.2$\pm$0.2  &   520$\pm$80  & \\ 
~H$\alpha_B$ & 6562$\pm$2  &    2400$\pm$200  & -70$\pm$80 \\ 
\hline 
\end{tabular}
\end{table}

In comparison with other BLRG, the nucleus of PKS~1932-464 has a relatively
weak radio core ($R_{8.6GHz}=S_{\rm core}/S_{\rm ext} < 0.006$, see also
Morganti {\it et al.} 1997), although its X-ray luminosity 
($L_x = 6\times10^{43}$, Siebert {\it et al.} 1996) is comparable
with other BLRG in the sample of southern 2Jy radio sources (Tadhunter {\it et al.} 1993).

\section{Discussion}

\subsection{The optical  morphology and the connection with the
radio structures}

The optical morphology shows complicated structures of ionized gas, 
including narrow
filaments and arcs.  Some of these structures may represent the remains of tidal
tails resulting from interactions with other galaxies, while other features
could be the result of interactions between the radio structures and the
ambient gas.  It is unlikely, however, that all the EELR are currently
interacting with the radio plasma, since the filament to the south lies well
away from the radio axis, and the arc-like feature to the east lies outside the
eastern radio lobe.  

As mentioned in section \S3.1 the main body of the galaxy shows double
structure in [OII]$\lambda$3727.  It is inte\-resting to note that a complex morphology in the central
nuclear region has been found in other radio galaxies, including 3C171 (Clark
{\it et al.} 1997b), and many of the targets observed during the HST Snapshot
Survey of 3CR radio galaxies (Koff {\it et al.} 1996).  In some cases it is
clearly due to the presence of an obscuring dust band.  The fact that
the double morphology is not apparent in the optical continuum image of PKS~1932-464
(Fig.~1, bottom-left) suggests that a) the structure is not due to dust, but rather due to a lack of emission line material or b) the continuum emission
is concentrated in a more compact region in such a way that the dust band
does not obscure it. In such a situation, the dark band would be apparent
in the line+continuum image, but not in the pure continuum image.  

Another possibility is that  the radio jet in its passed
through the extended gas has hollowed a channel 
({\it eg.} Jackson {\it et al.}, 1993)).  However, the direction of this cha\-nnel in PKS~1932-464 is rotated with respect
to the radio axis and this interpretation appears unlikely (see Fig.~1,
bottom-right).  Alternative possibilities will be discussed later. 

The morphology of feature ``C'' in Fig.~1 (the arc-like shape structure) is
reminiscent of the radio galaxy PKS~2250-41, where the western arc of line
emission lies just beyond the western radio lobe.  There are reasons to believe
that in PKS~2250-41 the W radio lobe is confined and constrainted by the line
emitting gas (see detailed discussion in Clark {\it et al.} 1997a).  In
PKS~1932-464, the physical connection is not so clear. From  the overlap of 
the radio and optical
images,  the hot spot lies $\sim$4 arc sec ($\sim$20kpc) from the inner part of
the arc.    Considering possible errors in astrometry, if we assume that the
radio core and the optical nucleus are at the same location, this distance 
is $\sim$14 kpc in projection.  There is also some offset in between the arc and radio
hotspot in PKS~2250-41, the offset distance is smaller in this object
($\sim$10 kpc).  Ano\-ther example is provided by Cyg A.  Carilli {\it et al.}
(1994)  detected an arc of discontinuous RM roughly concentric with a hotspot.  This might be a bow shock formed after the material is 
compressed by the shocks produced by
the advancing hot spot.   The
projected distance from the bow shock to the hot spot is $\sim$4.5 kpc. 

The arc in PKS~1932-464  presents some differences with the arc in PKS~2250-41
which make  less clear its physical association with the radio lobe:
 the line widths are
smaller  and the  ionization state much higher.   

\subsection{PKS~1932-464 and the high redshift radio galaxies.}

Despite its relatively low redshift, PKS~1932-464 shares a number of
characteristics with high $z$ ($z>0.7$) radio galaxies.  In particular,
McCarthy \& van Breugel (1989) have shown   the arm-length ratio (ratio
between the distance of the two lobes from the core) for 3CR radio gala\-xies is
strongly a function of redshift with the arm-length ratio increasing with
redshift.
The arm-length ratio observed in PKS~1932-464 is 1.56 that is
typical of objects with $z>0.7$ while at the redshift of PKS~1932-464 the
typical ratio is 1.18.  Moreover in high-z radio galaxies the extended
emission lines are brighter on the side of the radio lobe closest to the
nucleus (McCarthy \& van Breugel 1989, Pedelty {\it et al.}  1989) and this lobe is
also systematically more depolarized (Pedelty {\it et al.}  1989, Liu \& Pooley
1991).  These characteristics are also observed in PKS~1932-464.  The
arm-ratio asymmetry could have di\-fferent causes but the most likely is that
the propagation of the plasma jet is affected by an external medium
which is inhomogeneously and asymmetrically distributed around the galaxy (McCarthy \& van
Breugel 1989, Pedelty {\it et al.}  1989)
This scenario can also explain the higher depola\-rization in the
lobe closer to the nucleus: this lobe would be the one more deeply embedded in
the halo and therefore more highly depolarized. 

The distribution of EELR in PKS~1932-464 is
not such to explain the overall depolarization of the E lobe (i.e.  they
``cover'' only part of the radio lobe). This supports that an external 
medium  is responsible. If the warm gas (the gas in the EELR) traces the
distribution of hot external ISM, we expect a denser (asymmetric) medium on the E
side. This could  explain the higher depolarization.  This medium
could be the hot halo mentioned above.  Ano\-ther possibility is that 
the hot ISM  is
distributed in a denser sheath around the lobes as proposed in the case of
Cygnus~A by Dreher {\it et al.} (1987). 

We cannot exclude, however, that the EELR can still affect the depolarization
on smaller scales (Pedelty et al.  1989) and explain some of the patchy
distribution observed in the depolarization in PKS~1932-464.  In PKS~1932-464
we note that the region of higher depola\-rization in the eastern lobe is
coincident with one bright region of ionized gas, region B.

The radio sources in high-z objects are thought to show a parti\-cularly strong
interaction between the radio plasma and the environment ({\it e.g.} McCarthy {\it et al.} 1987, Best {\it et al.} 1996).  The fact that PKS~1932-464 as well as PKS~2250-41 (Clark {\it et
al.} 1997a) look so similar to these high-z objects suggests that there are systems at low/intermediate redshift where we see
situations {\sl very} similar to the high redshift systems.  The similarity
must be in the environment.  Together with other peculiarities, high-z radio
galaxies show an excess of companion galaxies detected along the axes of the
radio sources (R\"ottgering {\it et al.}  1996).  Inte\-restingly, both PKS~2250-41
and (possibly) PKS~1932-464 have  a companion.

\subsection{The line ratios and the emission line mechanism.}

The two  main  mechanisms which might be involved in the emission
line processes are: a) photoionization by the hard UV continuum emitted by the
central AGN, and b) high velocity radiative shocks which can influence the emission line processes due to  the generation of a strong local 
UV photon field in the hot post-shock zone,
which can ionize the surrounding medium both upstream and downstream. Moreover,
the radiative cooling of gas behind the hot post-shock zone can also give rise
to line emission. 

	The optical line ratios of PKS~1932-464 locate this object in
the general trend defined by low redshift radio galaxies in the diagnostic diagrams
involving optical lines (Robinson et al. 1987). However, the optical emission lines do not allow the discrimination between  shock  and AGN photoionization  
because the models overlap in
their predictions of the optical line ratios.

	In section \S3.5 we showed that the ionization level of the gas
decreases outwards from the nucleus across the inner EELR and rises
to high values in the outer EELR (see Fig.~5).

{\it Can this behaviour be explained in terms of pure AGN photoioni\-zation?}
The decrease in the ionization level across the inner EELR 
is consistent with AGN photoionization: the AGN continuum is diluted at 
increasing distances from the nucleus (varying as $\sim$1/$r^2$). For a 
gas density which decreases less steeply than $\sim$1/$r^2$  
we expect the ionization level
to decrease because of the steeper drop in the number of ionizing photons
compared to the number of atoms to be ionized.

However, the outer EELR shows a noticeably higher ionization level.
In the frame work of pure AGN photoionization, this implies that the density
must drop more rapidly that $\sim$1/$r^2$ between the inner and the
outer EELR. 
Another possibility is that projection effects mislead us. It is possible 
that  what
we have named inner EELR, is, indeed, more distant to the nucleus
than the one we have called outer EELR. In this case, the high level 
of ionization observed in the outer EELR could be due to the higher
density of ionizing photons. However, the projection factors would have
to be extreme for this explanation to hold.

\vspace{0.2cm}

{\it Can this behaviour be explained in terms of processes associated with
shocks?}
	The [OIII]/H$\beta$ ($\sim$3-5) and [OII]/[OIII] ($\sim$1.5-2) ratios  
in the inner EELR indicate a very low ionization level. Such values   are in 
good agreement with  those observed in radio galaxies with jet-cloud interactions (Clark 1996). The low
ionization state of the inner EELR is consistent with the compression
effect of the jet shocks. It is interesting to note that in other objects
with jet-cloud interactions a minimum in the ionization state is observed to be 
coincident with or just beyond the radio hot spots (for instance, 3C171, 4C29.30, ComaA: Clark 1997). In PKS~1932-464, this ionization minimum occurs well behind the
radio hot spot, close to  the inner edge of the E radio lobe. 
It is important to remember that  the slit did not cross the radio hot spot
where the main signs of interaction will occur. Therefore, we do not know whether  
the ionization level is even lower at this position. The low ionization level of the inner
EELR
suggests, independently of the  mechanism (shocks or outflowing wind), that
the gas has been compressed in this region. 

 As the gas in the outer EELR lies outside the radio source it is reasonable to think that it is 
ionized by the continuum emitted by the central AGN and/or by the strong UV continuum emitted by fast shocks. The [OIII]/H$\beta$  ($\sim$10) and [OII]/[OIII] ($\sim$0.4) emission line ratios support this idea: they indicate a high ionization level which cannot be explained 
in terms of shocked gas which is cooling down,
but rather gas which has not entered the shock and is being photoionized
by a strong UV conti\-nuum.

\subsection{The kinematics}

	As we mentioned before, the kinematical properties of the gas reveal
the presence of two well differentiated EELR regions, coincident with the
regions defined by the spatial line profiles (the inner and the outer EELR). 
Particularly interesting is the inner EELR.  It presents split components and
large line widths (FWHM$\sim$450 km s$^{-1}$) which suggest that its
kinematics has been perturbed.  Line widths of this order are not as extreme
as observed in some radio galaxies with jet-cloud interactions, such as 3C171
(Clark {\it et al.} 1997b), for which broader components (FWHM$\sim$1000 km
s$^{-1}$) are detected.    However, it is  interesting to remark that
the lines are noticeably broader than those emitted in the outer EELR
(FWHM$\sim$200 km s$^{-1}$). 

The fact that we do not detect extreme motions in the extended gas of
PKS~1932-464 does not mean that the kinematics is not highly perturbed.  The
slit was not exactly aligned with the radio axis, where the most extreme
motions are expected if the jet is interacting with the ambient gas. 
Moreover, it is possible that faster motions could be taking place
perpendicular to the line of sight.

Other radio galaxies with jet-cloud interactions  show a clear anticorrelation between line widths and the ioni\-zation state of the emi\-tting ion.  Low ionization lines are broader than high ionization lines (Clark et al. 1997b). This is observed
also in supernova remnants and is attributed to the fact that low ionization  emission lines
are produced mainly in the shocked, compressed  and accele\-rated gas, while the high emission
lines are mainly a contribution of the non-shocked gas (Greidanus \& Strom, 1992).
The spatial distribution of the velocity shifts and line widths does not reveal this behaviour
(Fig.~7) along PA270 in PKS~1932-464.
  We have checked if
the spatially integrated spectra for the different spatial regions  (outer EELR, and components A
and B of the inner EELR) show this same behaviour. We found some differences among the lines
emitted in component B of the inner EELR, with the low ionization lines being broader than the
high ionization lines. This effect, however, is likely  to be due to a velocity gradient across
the gaseous region.

\section{PKS~1932-464: a jet-cloud interaction in a radio galaxy?}

Clark {\it et al.} have studied a small sample of intermedia\-te redshift radio
galaxies which were already known to show
clear evidence for jet-cloud interactions. This evidence comes both
from spectroscopy and imaging. The targets show some special properties
which are a consequence of the interaction between the radio jet
and the am\-bient gas in the galaxy. Some of these properties have already
been mentioned. 

	a) Striking correspondence between line emission and radio structures

	b) Highly perturbed kinematics shown by complex emission line profiles: split narrower components ($\Delta v\sim$550km s$^{-1}$) and an underlying broad component (FWHM$\sim$1000 km s$^{-1}$) are often observed.

	c) Anti-correlation between line width and ionization state.

	d) Association of ionization minima with shocked structures. The compression effect and maybe also shock ionization produce a drop on the
ionization level of the gas.

	{\it Do we see any of these properties in PKS~1932-464?}

	a) The overlay of the radio and optical maps show that the 
outer EELR lies  outside the radio structure and appears to circumscribe
the E radio lobe. 
However, the large offset between
the radio hotspot and the outer arc, the  apparently quiescent 
kinematics of the emi\-tting
gas in this region (with narrow lines and simple profiles), and the high level
of ionization, do not provide  evidence
 for such an interaction and the
observed properties of this gas could well be explained in terms of AGN
illumination. The inner EELR lies well inside the radio structures and an
interaction is plausible. 
  
	On the other hand, PKS~1932-464 shares many similarities with high
redshift (z$>$0.7) radio galaxies: the radio lobe arm-ratio, radio depolarization and  optical asymmetries. 
These similarities suggest that, as in  high
redshift radio galaxies,  PKS~1932-464 lies in a rich environment which is
interacting with the radio jet.

	b) Our study of the kinematics of the EELR does not show extreme
motions along PA270. 
However, the differences in the kinematic
properties of the outer and inner EELR suggest that different mechanisms are
influencing the kinematics of the two regions: the gas in the inner EELR could
be affected by jet-induced shocks.

	c) There is  not clear evidence for anti-correlation between the widths of 
the lines and the ionization level of the original ion as due to the effects of shocks.

	d) We find that the level of ionization of the inner EELR is noticeably
lower than the level of ionization of the outer EELR and similar to that measured in other jet-cloud interaction targets. Interestingly 
the inner EELR shows also the broader lines.

Therefore, although PKS~1932-464 does not show compelling evi\-dence for 
jet-cloud interactions, some of its characteristics suggest that, indeed, the
radio jet is interacting with the ambient gas in the inner EELR. Obtaining
spectra along the radio axis, where the strongest effects of the 
interaction  are expected, will provide more definitive answers.

\subsection{Alternative possibilities}

Can we explain the properties of PKS~1932-464 in another way which does not
involve the interaction between the radio jet and the ambient gas?

{\it An expanding shell}

	The inner EELR corresponds, as we said before, to the feature in 
Fig.~1, which we named 'ARM'. This arm seems to have a faint extension
which closes into a loop connected to the main body of the galaxy. Such morpho\-logy
is reminiscent of an expanding shell which could be due, for instance to an
outflowing wind. If the wind is sweeping material out, this could explain the morphology described in section \S3.1 , where a gap is seen 
between the arm and the main
body of the galaxy. 

An observable characteristic of outflowing winds is 
broadening of the lines ({\it e.g.} Legrand et al 1997) as is observed
in the long wavelength component. Component B (the one more distant to the
nucleus) could be more
distant gas which has
not  yet been reached by the wind and therefore emits narrower lines.
The different velocity shifts for the two components support this.
The fact that the
velocity shifts are smaller for  component A
(supposedly, the one expanding)
could be explained if  the direction of
the expansion takes place close to the plane of the sky.

	If the wind compresses the gas, the density enhancement would explain
the low level of ionization observed for the gas in this region. We should
explain, however, why  component B shows also such low ionization level,
if it has not been reached by the expanding shell.

{\it Mergers or interactions between two galaxies}

	The projected thickness of the inner EELR (the 'ARM')  is $\sim$12 kpc and length $\sim$30
kpc which could be a smaller galaxy which is cannibalized or being merged with
PKS~1932-464. Interaction between galaxies can also disturb
the kinematics of the ionized gas and the emission lines appear broad
and/or with split components. Such is the case of  the radio galaxy 
 PKS0349-27 (Koekemoer \& Bicknell 1994 ).

\section{Summary and Conclusions}

Although PKS~1932-464 does not show compelling evidence for 
jet-cloud interactions, some of its characteristics suggest that such
an interaction might be taking place.
 
We do not find the extreme properties observed in 
other jet-cloud interaction radio galaxies (striking correspondence between
radio and optical structures, extreme motions, conclusive anti-correlation between line width and ionization level).
However, this object presents
two well differentiated EELR with very different kinematics and ionization
level of the gas. The broader lines and split components in the inner EELR, 
its low ionization level and the fact that it lies well inside the radio
structures could be explained in terms of shocks produced during the
interactions between the radio structures and the emitting gas.
On the other hand, the narrow and single line profiles, the high ionization
level of the gas and the fact that it lies outside the radio structures,
suggest that the properties of the outer EELR are explained in terms
of pure AGN illumination. It is also possible  that this region is
 the  precursor gas ahead of the shock and photoionized
by its UV continuum.

	PKS~1932-464 shares many similarities with high redshift  radio galaxies:
the radio lobe arm-ratio, radio depolarization and  optical asymmetries.
This suggests that, as in  high
redshift radio gala\-xies, the radio plasma is  interacting strongly with 
the probably rich environment of PKS~1932-464.

	We have considered two alternative scenarios to explain the 
observed properties of PKS~1932-464. One of them involves an outflowing wind 
which is forcing the 
gas in the inner EELR to expand, producing the broadening of the lines
and the compression of the gas. It could also be that the inner EELR
is a companion galaxy which is interacting with PKS~1932-464.

	The spectroscopic information was obtained with a slit which was not
exactly aligned along the radio axis. If the radio jet is interacting with the
ambient gas we expect the strongest disturbances to take place near
this direction.
Obtaining long slit spectra along the radio axis can provide
more conclusive answers. 

 The nuclear spectrum reveals the presence of a broad component in H$\alpha$
which indicates that PKS~1932-464 is a broad line radio galaxy in
which the AGN is observed directly. This 
explains 
the low
level of polarization observed in the UV continuum of this object.
 
\vspace{1cm}

{\Large \bf Acknowledgements} 

\vspace{0.7cm}
 M.Villar-Mart\'\i n thanks
Jacco van Loon for useful discussions.  M.Villar-Mart\'\i n acknowledges 
support from PPARC grant. Thanks also to the referee for his/her useful comments.
\newpage

{\Large \bf Bibliography} 

\normalsize
\vspace{0.7cm}

 Baum S.A., Heckman T., van Breugel W., 1990,
ApJS, 74, 389

 Best P.N., Longair M.S., R\"ottgering H.J.A.,
1996, MNRAS, 280, L9

  Binette L., Wang J.C.L., Zuo L., Magris C.M., 1993a, AJ, 105, 797
 
 Binette L.,  Wang J.C.L., Villar-Mart\'\i n M.,
Martin P.G., Magris C.M., 1993b, ApJ, 414, 535

 Binette L., Wilson A.S., Storchi-Bergman T., 
1996, A\&A, 312, 365

 Burstein D., Heiles C., 1984, ApJS, 54, 33-79
 
 Cardelli J., Clayton G., Mathis J., 1989, ApJ, 345, 245

 Carilli C.L., Perley R.A., Dreher. J.H., 1988, ApJ, 334, 73

 Chambers K.C., Miley G.K., van Breugel W.J.M., 1987, Nature, 329, 604

 Clark N.E. \& Tadhunter C.N. 1996, in {\em Cygnus A-- Study of a 
Radio Galaxy} (eds. Carilli C.L. \& Harris D.E.), CUP, p15
 
Clark N.E., 1996, Phd Thesis, University of Sheffied 

 Clark N.E., Tadhunter C.N., Morganti R., Killeen N.B., Fosbury 
R.A.E., Hook R.N., Shaw M., 1997a, MNRAS, 286, 558

 Clark N.E., Axon D., Tadhunter C.N., Robinson A., O'brien P., 1997b, ApJ, in press
 
  Dickson R., 1997,  Phd Thesis, University of Sheffied 

 Dopita M.A., Sutherland R.S., 1995, ApJ, 455, 468

 Dopita M.A., Sutherland R.S., 1996, ApJS, 102, 161

 Dreher J.W., Carilli C.L., Perley R.A., 1987,
ApJ, 316, 611

 Garrington S.T., Conway R.G., 1991, MNRAS,
250, 198 

 Greidanus H., Strom R.G., 1992, A\&A, 257, 265

 Heckman T.M., Illingworth G.D., Miley G.K., van Breugel W.J.M., 1985, ApJ, 299, 41
 
Jackson N., Sparks W.B., Miley G.K., Machetto F., 1993, A\&A, 269, 128

 Koekemoer A.M., Bicknell G.V.,  1994, in {\it Extragalactic Radio Sources}, IAU 175, p. 473

 de Koff S., Baum S., Sparks W., Biretta J., Golombek D., Macchetto F., McCarthy P., Miley G., 1996, ApJS, 1996, 107, 621 

 Legrand F., Kunth D., Mas-Hesse J.M., Lequeux J., 1997, A\&A, in press

 Liu R., Pooley G., 1991, MNRAS, 253, 669

 McCarthy P.J., Spinrad H., Djorgovsky S.,
Strauss M.A., van Breugel W., Liebert J., 1987, ApJ, 319, L39

 McCarthy P.J., van Breugel W., 1989, in {\em ESO Workshop on Extranuclear Activity in Galaxies }, eds. E. Meurs and R.A. Fosbury, 
ESO Conf. and Workshop Proc. no 32,  
p55

 McCarthy P.J., Baum S., Spinrad H., 1996, ApJS, 
106, 281

 Morganti R., Oosterloo T.A., Reynolds J.,
Tadhunter C.N., Migenes V., 1997, MNRAS, 284, 541

  Pedelty J.A., Rudnick L., McCarthy P.J.,
Spinrad H., 1989, AJ, 97, 647

 Pentericci L., Roettgering H., Miley G., 
Carilli C., McCarthy P., 1997, A\&A, 326, 580

 Robinson A., Binette L.,
Fosbury R.A.E., Tadhunter C.N., 1987, MNRAS 227, 97

R\"ottgering H.J.A., West M., Miley G.  \&
Chambers K.  1996, A\&A, 307, 376

 Siebert J., Brinkmann W., Morganti R., 
Tadhunter C.N., Danzinger I.J., Fosbury R.A.E., di Serego Alighieri S., 1996,
MNRAS, 279, 1331 

 Tadhunter C.N., Fosbury R.A.E., Quinn P.J., 1989, MNRAS, 240, 225

 Tadhunter C.N., 1990, in {\em New Windows to
the Universe}, eds. F. S\'anchez and M. V\'azquez, Cambridge University Press,
 p175

 Tadhunter C.N., Morganti R., di Serego 
Alighieri S., Fosbury R.A.E., Danzinger I.J., 1993, MNRAS, 263, 999

 van Ojik R., Roettgering H., Carilli C.L.,
Miley G.K., Bremer M.N., Macchetto F., 1996, A\&A, 313, 25

\end{document}